# Comparative Evaluation of Generative AI Models for Chest Radiograph Report Generation in the Emergency Department

Original Research

Woo Hyeon Lim, MD,[a,*] Ji Young Lee, MD, PhD,[a,*] Jong Hyuk Lee, MD, PhD,[a,b] Saehoon Kim, PhD,[c,†] Hyungjin Kim, MD, PhD[a,b,c,†]

[a]Department of Radiology, Seoul National University Hospital, 101 Daehak-ro, Jongno-gu, Seoul 03080, Korea; [b]Department of Radiology, Seoul National University College of Medicine, 101 Daehak-ro, Jongno-gu, Seoul 03080, Korea; [c]Soombit.ai, 21 Pangyo-ro 255-gil, Bundang-gu, Seongnam 13486, Korea

[*]These authors contributed equally to this study as co-first authors.

[†]These authors contributed equally to this study as co-senior authors.

**Address correspondence to:**

**Hyungjin Kim, MD, PhD.** Department of Radiology, Seoul National University College of Medicine, 101 Daehak-ro, Jongno-gu, Seoul, 03080, Korea. Tel: 82-2-2072-2254, Fax: 82-2-743-6385, E-mail: khj.snuh@gmail.com

## Abstract

**Background:** Medical domain vision–language models (VLMs) have recently been introduced for chest radiograph (CXR) report generation. Comparative evaluation is essential to assess their clinical applicability.

**Purpose:** To benchmark open-source or commercial medical image–specific VLMs against real-world radiologist-written reports, focusing on diagnostic quality, clinical acceptability, hallucinations, and language clarity.

**Methods:** This retrospective study included adult patients who presented to the emergency department of a tertiary center between January 2022 and April 2025 and underwent same-day CXR and CT for febrile or respiratory symptoms. Reports from five VLMs (AIRead, Lingshu, MAIRA-2, MedGemma, and MedVersa) and radiologist-written reports were randomly presented and blindly evaluated by three thoracic radiologists using four criteria: RADPEER, clinical acceptability, hallucination, and language clarity. Comparative performance was assessed using generalized linear mixed models, with radiologist-written reports treated as the reference. Finding-level analyses were also performed with CT as the reference.

**Results:** A total of 478 patients (median age, 67 years [interquartile range, 50–78]; 282 men [59.0%]) were included. AIRead demonstrated the lowest RADPEER 3b rate (5.3% [76/1434] vs. radiologists 13.9% [200/1434]; *P*<.001), whereas other VLMs showed higher disagreement rates (16.8–43.0%; *P*<.05). Clinical acceptability was the highest with AIRead (84.5% [1212/1434] vs. radiologists 74.3% [1065/1434]; *P*<.001), while other VLMs performed worse (41.1–71.4%; *P*<.05). Hallucinations were rare with AIRead, comparable to radiologists (0.3% [4/1425]) vs. 0.1% [1/1425]; *P*=.21), but frequent with other models (5.4–17.4%; *P*<.05). Language clarity was higher with AIRead (82.9% [1189/1434]), Lingshu (88.0% [1262/1434]), and MedVersa (88.4% [1268/1434]) compared with radiologists (78.1% [1120/1434]; *P*<.05). Finding-level analyses showed substantial variability in sensitivity across VLMs for the common findings (i.e., lung opacity, pleural effusion, cardiomegaly, and emphysema): AIRead,

15.5–86.7%; Lingshu, 2.4–86.7%; MAIRA-2, 6.0–72.0%; MedGemma, 4.8–76.7%; and MedVersa, 20.2–69.3%.

**Conclusion:** Medical VLMs for CXR report generation exhibited variable performance in report quality and diagnostic measures.

## Introduction

Over the past decade, artificial intelligence (AI)-based approaches, including deep learning, have shown promise in detecting and localizing abnormalities in chest radiographs (CXRs)[1]. More recently, the emergence of vision–language models (VLMs), which jointly process images and text, has brought attention to radiologic report generation[1]. This is particularly relevant in light of the global shortage of radiologists, which is increasingly strained by the growing demand and clinical importance of imaging studies[2,3,4]. VLM-based report generation has therefore been proposed as a potential solution for producing timely and clinically useful reports in settings with limited radiologist availability[1].

Previous studies have explored the performance of AI in CXR report generation, most often evaluating outputs with semantic similarity measures or automated metrics based on label extraction[5,6,7,8]. However, assessing the clinical utility and quality of AI-generated reports requires a multifaceted approach to accurately determine the readiness of VLMs for clinical translation[9,10]. For instance, one prior study employed multidimensional assessments including clinical acceptability, expert agreement, and overall report quality[10]. Key dimensions of VLM-generated report evaluation should encompass diagnostic performance (report-level and finding-level), clinical acceptability (i.e., whether generated reports are suitable for clinical use with only minor revisions), absence of hallucination, and linguistic clarity[10,11].

Medical image–specific VLMs can internalize radiology priors beyond what generalist models acquire from natural images and captions[10,11,12]. Although medical VLMs have demonstrated promising performance in isolation, the literature currently lacks a systematic, standardized comparison across multiple medical VLMs for the specific task of CXR report generation[13,14,15,16]. It is essential to compare model-generated reports under a unified framework and identical experimental conditions.

In this retrospective study, we address this gap by conducting a systematic head-to-head benchmarking of medical image–specific VLMs for CXR report generation. We performed a multidimensional comparison of five recently released models against real-world radiologist-written reference reports, using CXRs obtained from patients presenting to the emergency department.

## Materials and Methods

This retrospective study was approved by the Institutional Review Board of Seoul National University Hospital (No. H-2505-199-1646), with a waiver of informed consent.

### Patients

Adult patients presenting to the emergency department of a tertiary referral center between January 2022 and April 2025 were retrospectively identified. Eligible patients were those who underwent initial frontal CXR (anteroposterior or posteroanterior view) and chest CT on the same day, with CT serving as the reference standard for finding-level assessment. Among these, patients presenting with fever and/or chills or respiratory symptoms—including dyspnea, cough with or without sputum, blood-tinged sputum or hemoptysis, and chest discomfort or pain—were included. The following patients were excluded: 1) those whose CXR reports contained comparison statements, 2) those whose reports were not available, and 3) those whose reports relied heavily on clinical information rather than imaging findings.

### Data Collection

From the electronic medical records, we collected patient demographics, radiologist-written CXR reports, the interval between CXR and CT scan, chief complaints, and comorbidities. Digital Imaging and Communications in Medicine files of CXRs were exported from the Picture Archiving and Communication System. CXRs between January 2022 and February 2024 were primarily interpreted by radiology residents, whereas those between March 2024 and April 2025 were interpreted by board-certified radiologists. The latter period coincided with a nationwide strike by medical students, residents and fellows, which shifted the emergency department workload exclusively to board-certified radiologists[17,18].

## CXR Report Generation

Five medical image–specific VLMs were employed: AIRead (Soombit.ai, Korea), Lingshu (Alibaba, China), MAIRA-2 (Microsoft Research, UK), MedGemma (Google DeepMind, UK), and MedVersa (Harvard University, MA)[10,13,14,15,16]. Among these, the latter four models are open-source. For each model, CXR reports were generated from the corresponding Digital Imaging and Communications in Medicine files, without access to clinical information or comparison images. Model inference was performed on a single NVIDIA H100 graphics processing unit running CUDA 12.4, without fine-tuning or transfer learning. Details of each model including their availability are provided in Supplementary Text 1.

## Evaluation of VLM-generated CXR Reports

VLM-generated CXR reports were evaluated using two approaches: 1) a multireader study assessing the diagnostic quality of free-text reports based on four criteria (RADPEER score, clinical acceptability, presence of hallucinations, and language clarity)[10,19], and 2) finding-level diagnostic sensitivity and specificity.

### *Report-level Diagnostic Quality*

The RADPEER scoring system was originally developed by the American College of Radiology as a peer review tool for quality assurance[19]. It uses a three-point scale: a score of 1 indicates complete agreement with the original interpretation, while scores of 2 (understandable miss) and 3 (unacceptable miss) indicate increasing levels of disagreement. Subcategories “a” and “b” further classify findings as clinically insignificant or clinically significant, respectively. The study outcomes were defined as the proportions of reports with a score of 3b (unacceptable, clinically significant miss) and with scores of 2b or 3b combined (any clinically significant miss [i.e., stricter criterion]).

Clinical acceptability was evaluated on a four-point scale that considered the presence of detection errors (e.g., false negatives, false positives, incorrect laterality) and the extent of editing required for clinical use, complementing the RADPEER system[11]. In this scale, a score of 1 denoted a report that was not acceptable because of detection errors for clinically referable abnormalities. A score of 2 denoted a report acceptable after major revision due to less critical detection errors or the need for extensive edits. A score of 3 denoted a report acceptable only with minor revision, being generally accurate but requiring limited edits. Finally, a score of 4 denoted a report acceptable as is, being accurate, clear, and clinically actionable without edits. The study outcome was the proportion of clinically acceptable reports, defined under two thresholds: 1) the standard criterion, encompassing reports acceptable as is or after minor revision, and 2) the stringent criterion, limited to reports acceptable as is.

Hallucination was defined as the inclusion of information that could not be derived from a single frontal CXR. Examples included references to a lateral view, patient history, comparison studies, or measurements (such as lesion size or distance) not supported by the study setting and models. Language clarity was evaluated as a measure of readability using a five-point scale (1, unacceptable; 2, poor; 3, moderate; 4, good; 5, excellent). The study outcome was the proportion of reports rated 4 or 5.

This evaluation framework was applied to reports generated by the five VLMs as well as to radiologist-written reports. Three board-certified, fellowship-trained thoracic radiologists (J.H.L., 4 years of post-fellowship experience; W.H.L., 2 years; J.Y.L., 1 year) independently evaluated all reports using a web-based assessment platform in a randomized, blinded manner. The criteria are summarized in Supplementary Table 1.

***Per-patient Finding–level Analyses***

Finding-level sensitivity and specificity were measured for both VLM-generated and radiologist-written CXR reports, with same-day chest CT as the reference. CT scans were reviewed for the presence of predefined abnormalities: aortic aneurysm (≥50 mm for the ascending aorta, ≥40 mm for the descending aorta), cardiomegaly (CT cardiothoracic ratio >0.5), cavity, emphysema, hilar mass, lung opacity (consolidation or ground-glass opacity), mediastinal mass, miliary nodules, non-calcified nodule (≥4 mm) or mass, pericardial effusion, pleural effusion, pneumomediastinum, pneumoperitoneum, pneumothorax, and reticular opacity[20,21,22]. Each CT was reviewed by one of three board-certified thoracic radiologists after completion of the reader study.

Then, a previously developed labeler was used to extract the presence or absence of individual findings from the CXR reports (Supplementary Text 2). Briefly, CheXGPT, a Bidirectional Encoder Representations from Transformers-based model trained on 3 million CXR reports was used[23].

**Statistical Analyses**

The normality of continuous variables was assessed using the Shapiro–Wilk test. Continuous variables were presented as mean ± standard deviation or as median with interquartile range (IQR), depending on distribution.

The proportions of report quality measures were analyzed using generalized linear mixed-effects models with a logit link. Report source (VLMs or radiologists) was included as a fixed effect, using radiologists as the reference category, while reader and case were modeled as random intercepts. Cases in which original reports contained measurements were excluded from the hallucination analysis. Subgroup analyses were performed, stratified by sex (men vs. women), age (≥65 vs. <65 years), and the expertise level of the interpreting radiologist (resident vs. board-certified). For finding-level analyses, sensitivity and specificity with 95% confidence intervals were calculated for each report source and for each abnormality, using CT as the reference standard.

All statistical analyses were performed using R (version 4.1.2) by W.H.L., and a two-sided *P* value <.05 was considered statistically significant.

## Results

### Patients

Of the 1037 adult patients who presented to the emergency department and underwent initial CXR and CT, the following were excluded: 1) patients without fever, chills, or respiratory symptoms (n=537), 2) patients whose CXR reports contained comparison statements (n=19), 3) patients with missing reports (n=1), and 4) patients whose reports relied heavily on clinical information rather than imaging findings (n=2). A total of 478 patients (median age, 67 years [IQR, 50–78]; 282 men [59.0%]) were included in the final analyses (Fig. 1).

The most common chief complaint was dyspnea (47.7% [228/478]), followed by fever and/or chills (23.6% [113/478]) (Table 1). Malignancy was present in 5.4% (26/478) of patients (lung cancer, n=9; other solid organ malignancy, n=12; hematologic malignancy, n=5), and diffuse parenchymal lung disease was identified in 3.1% (15/478; chronic obstructive pulmonary disease, n=10; interstitial lung disease, n=5) (Table 1).

### Report-level Diagnostic Quality

The median word count was 5 words (IQR, 4–7) for radiologists, 10 (IQR, 4–15) for AIRead, 32 (IQR, 27–38) for Lingshu, 19 (IQR, 14–28) for MAIRA-2, 275 (IQR, 242–321) for MedGemma, and 46 (IQR, 32–66) for MedVersa.

The proportion of radiologist-written reports containing RADPEER 3b disagreement was 13.9% (200/1434). Among the VLM-generated reports, compared with radiologists' reports, AIRead showed a lower proportion of RADPEER 3b (5.3% [76/1434], *P*<.001), whereas other VLMs showed higher proportions: Lingshu (43.0% [616/1434], *P*<.001), MAIRA-2 (24.5% [352/1434], *P*<.001), MedGemma (16.8% [241/1434], *P*=.02), and MedVersa (25.9% [371/1434], *P*<.001) (Table 2, Fig. 2A). For RADPEER 2b or 3b disagreement, the proportion was also lower with AIRead compared with radiologists (AIRead, 20.3% [291/1434] vs. radiologists, 30.3% [435/1434]; *P*<.001), but higher with

Lingshu (57.1% [819/1434], *P*<.001), MAIRA-2 (36.8% [528/1434], *P*<.001), and MedVersa (38.4% [551/1434], *P*<.001) (Table 2, Fig. 2A). MedGemma demonstrated RADPEER 2b or 3b rates (27.9% [400/1434], *P*=.10) comparable to those of radiologists.

The proportion of clinically acceptable reports using the standard criterion differed among the VLMs. AIRead showed the highest acceptability, higher than that of radiologists (AIRead, 84.5% [1212/1434] vs. radiologists, 74.3% [1065/1434]; *P*<.001) (Table 2, Fig. 2B). All other models showed lower acceptability than radiologists: Lingshu (41.1% [589/1434]), MAIRA-2 (65.6% [941/1434]), MedGemma (71.4% [1024/1434]), and MedVersa (63.7% [914/1434]) (*P*<.05 for all) (Table 2, Fig. 2B). Under the stringent criterion, AIRead still showed the highest acceptability, which was greater than that of radiologists (58.2% [834/1434] vs. 43.5% [624/1434], *P*<.001) (Table 2, Fig. 2B). MedGemma demonstrated an acceptability rate similar to that of radiologists (42.9% [615/1434], *P*=.68), whereas the other VLM-generated reports showed lower acceptability (25.3–39.7%, *P*<.05 for all).

A single hallucination was identified in the radiologist-written reports, which was not a true hallucination but rather a reviewer's subjective interpretation of a false-positive finding. The proportion of reports containing hallucinations was considerably higher in Lingshu (11.0% [157/1425]), MAIRA-2 (17.4% [248/1425]), MedGemma (5.4% [77/1425]), and MedVersa (12.3% [175/1425]) than in radiologists (0.1% [1/1425], *P*<.05 for all) (Table 2). In contrast, hallucinations were markedly suppressed in AIRead, with only 0.3% of reports (4/1425, *P*=.21) affected.

In terms of language clarity, AIRead (82.9% [1189/1434], *P*=.001), Lingshu (88.0% [1262/1434], *P*<.001), and MedVersa (88.4% [1268/1434], *P*<.001) outperformed radiologists (78.1% [1120/1434]). MedGemma had the lowest proportion of clear language among report sources (69.7% [1000/1434], *P*<.001) (Table 2, Fig. 2C).

These trends remained similar in subgroups stratified by sex, age, and the experience level of the interpreting radiologists (Supplementary Tables 2–7). Representative cases are shown in Figs. 3–5.

**Finding-level Diagnostic Performance**

The most frequent CT findings were lung opacity (consolidation or ground-glass opacity) in 53.8% (257/478), followed by pleural effusion in 42.3% (202/478), nodule or mass in 23.4% (112/478), emphysema in 18.2% (84/478), and cardiomegaly in 15.7% (75/478) (Tables 3–4).

Per-patient sensitivity and specificity for detecting major abnormalities potentially related to patients' acute symptoms (i.e., lung opacity and pleural effusion) varied across VLMs. AIRead and MedGemma achieved the highest sensitivity for lung opacity and pleural effusion, respectively (AIRead for lung opacity: sensitivity, 77.8% [200/257], specificity, 70.6% [156/221]; MedGemma for pleural effusion: sensitivity, 71.8% [145/202], specificity, 75.0% [207/276]). Lingshu showed the lowest sensitivities for both abnormalities among models (lung opacity: sensitivity, 16.3% [42/257], specificity, 89.1% [197/221]; pleural effusion: sensitivity, 21.3% [43/202], specificity, 90.2% [249/276]). VLMs also exhibited substantial variability for other findings (cardiomegaly: sensitivity, 69.3–86.7% and specificity, 61.5–87.3%; emphysema: sensitivity, 2.4–20.2% and specificity, 94.7–99.2%) (Tables 3–4, Fig. 6).

None of the VLMs identified any of the 10 cases of miliary nodules or the single case of pneumoperitoneum (Tables 3–4, Fig. 6). Lingshu failed to detect any of the nine pneumothorax cases (Table 3, Fig. 6), and AIRead successfully detected the one case of pneumomediastinum (Fig. 5). Additional results are presented in Tables 3–4 and Fig. 6.

## Discussion

Although several VLMs for medical imaging have recently been released, comparative evaluations remain limited. In this study, we analyzed CXRs from patients presenting to the emergency department with febrile or respiratory symptoms and compared the performance of five VLMs for report generation against radiologist-written reports acquired from clinical practice. Among the models, AIRead consistently demonstrated the most favorable balance across multiple quality dimensions, achieving lower rates of clinically significant disagreement and higher clinical acceptability, outperforming radiologists. The other VLMs exhibited variable results, with some excelling in language clarity but performing less reliably in diagnostic agreement and/or clinical acceptability. Notably, AIRead demonstrated very low hallucination rates, a desirable feature that enhances report reliability.

A key finding of our work is the discrepancy between diagnostic performance and readability. Fluent, well-structured text does not guarantee clinically accurate or useful content, highlighting that linguistic clarity and diagnostic fidelity are separate dimensions. That is, high language clarity, as observed in Lingshu and MedVersa, did not correspond to greater diagnostic accuracy, whereas MedGemma demonstrated the reverse pattern. MedGemma's exhaustive and enumerative reporting style performed well under the RADPEER scoring system, which primarily emphasizes report-level agreement; however, this approach appears to compromise readability, thereby limiting its practical utility.

Given that AIRead demonstrated lower rates of clinically significant disagreement (RADPEER 3b), higher acceptability under the stringent criterion, and rare hallucinations, the model may serve as a valuable adjunct in clinical practice. One potential use case is automated draft generation, in which AIRead produces a preliminary report that can be rapidly reviewed and finalized by radiologists, thereby improving efficiency without compromising diagnostic quality. In high-volume or resource-limited settings, AIRead could also function as a second reader to reduce error rates and enhance report consistency. Furthermore, its low hallucination rate suggests potential applicability in emergency or frontline care, where rapid and reliable reporting is essential. These indications underscore the role of such models not as replacements but as clinical support tools that augment radiologist performance and help maintain reporting quality under heavy workload conditions.

At the finding level, VLMs demonstrated relatively high sensitivities for common abnormalities, including lung opacity, pleural effusion, nodules or masses, and cardiomegaly. However, the reduced specificity observed among some models underscores a trade-off that may lead to false-positive downstream investigations. More critically, both radiologists and VLMs underperformed in detecting rare but clinically relevant findings, such as aortic aneurysms, mediastinal or hilar abnormalities, and pneumoperitoneum. Delayed recognition of these conditions may lead to catastrophic consequences. Therefore, further refinement of VLMs to better identify rare, high-impact abnormalities is essential. It is also important to note that the reference standard for each finding was based on CT results, and the inherent limitations of CXRs in detecting certain abnormalities—such as emphysema, miliary nodules, and pneumoperitoneum—should be considered when interpreting our results, as sensitivities may have been underestimated under these reference standards.

This study has several limitations. First, it was conducted retrospectively at a single institution. Second, CXR reports in Korea are typically written in a concise format, which may limit the generalizability of our findings to countries where more descriptive reporting is standard practice. Third, the diagnostic performance of radiologists in our study may have been underestimated. During the study period, a nationwide strike by medical students, residents, and fellows increased the workload of board-certified radiologists, who had to cover emergency department shifts in addition to their routine clinical duties. This unusual clinical environment may have negatively influenced their performance and should be considered when interpreting the results. Fourth, our evaluation included only one commercial model, with the remainder being open-source, as no other commercial medical image–specific VLMs were available during the study period. Fifth, this study did not assess the downstream impact on patient outcomes. Importantly, improved diagnostic performance does not necessarily translate into favorable clinical outcomes[24,25]. Sixth, the input to VLMs was limited to images alone, and their performance may be enhanced by incorporating relevant clinical metadata[26,27]. Finally, this study did not assess human–AI collaboration, which may further influence the clinical utility of VLM-generated reports.

In conclusion, a few VLMs for CXR report generation are now available, yet our evaluation demonstrated considerable variation in their performance. Among them, one commercial model not only outperformed radiologists in diagnostic quality but also showed an exceptionally low rate of hallucination, which may enhance trust in its outputs.

## Author Contributions

**W.H.L.** and **J.Y.L.** contributed equally to this work (co-first authorship), particularly in data collection, methodology, and investigation. **J.H.L.** contributed to this work in data collection, and investigation. **S.K.** and **H.K.** contributed equally as senior authors, with primary roles in conceptualization, methodology, resources, software, project administration, and supervision. All contributors have sufficiently participated and meet the criteria for authorship, taking full responsibility for the content of the study. All listed authors have approved the manuscript for submission.

## Conflicts of Interests

**W.H.L.** No relevant relationships.

**J.Y.L.** No relevant relationships.

**J.H.L.** received consulting fees from RadiSen; a research grant from Coreline Soft and Kakao Brain, outside of submitted work.

**S.K.** is the chief technical officer and a shareholder of Soombit.ai.

**H.K.** Research grants from Kakao Brain and RadiSen, an honorarium from AstraZeneca, consulting fees from RadiSen, stock and stock options in Medical IP, stock in Soombit.ai, and medical director of Soombit.ai.

**S.K.** is the Chief Technical Officer and a shareholder of Soombit.ai, and **H.K.** is a shareholder of Soombit.ai and serves as its Medical Director. All aspects of data control and analysis were independently performed by W.H.L., who has no conflicts of interest and no affiliation with Soombit.ai. S.K. and H.K. had no role in data management or analysis.

## Funding

Author declared no funding for this work.

## Data sharing

Data generated or analyzed during the study are available from the corresponding author by request.

# Tables

**Table 1:** Characteristics of Study Patients

| Variable | No. of Patients (n=478) |
|---|---|
| Age, years* | 67 (50–78) |
| Sex | |
| Men | 282 (59.0) |
| Women | 196 (41.0) |
| Chest CT-CXR interval, days* | 0 (0–0) |
| Projection | |
| Posterior-anterior | 237 (49.6) |
| Anterior-posterior | 241 (50.4) |
| Chief complaint | |
| Fever and/or chill | 113 (23.6) |
| Dyspnea | 228 (47.7) |
| Cough and/or sputum | 43 (9.0) |
| Blood-tinged sputum or hemoptysis | 32 (6.7) |
| Chest discomfort or pain | 62 (13.0) |
| Comorbidity | |
| Lung cancer | 9 (1.9) |
| Other solid organ malignancy | 12 (2.5) |
| Hematologic malignancy | 5 (1.0) |
| Chronic obstructive pulmonary disease | 10 (2.1) |
| Interstitial lung disease | 5 (1.0) |
| Heart failure | 12 (2.5) |
| Hypertension | 156 (32.6) |
| Diabetes mellitus | 94 (19.7) |
| Stroke | 15 (3.1) |
| Coronary artery disease | 17 (3.6) |

Data are presented as number of patients (percentage), unless otherwise specified.

*Data are presented as medians (interquartile range).

CT = computed tomography, CXR = chest radiograph.

**Table 2:** Evaluation of Radiologist-Written and AI-Generated Reports by Three Thoracic Radiologists

| Variable | Radiologist | AIRead | Lingshu | MAIRA-2 | MedGemma | MedVersa |
|---|---|---|---|---|---|---|
| RADPEER Score 3b, % | 13.9 (200/1434) | 5.3 (76/1434) | 43.0 (616/1434) | 24.5 (352/1434) | 16.8 (241/1434) | 25.9 (371/1434) |
| P Value | Reference | <.001 | <.001 | <.001 | .02 | <.001 |
| RADPEER Score 2b or 3b, % | 30.3 (435/1434) | 20.3 (291/1434) | 57.1 (819/1434) | 36.8 (528/1434) | 27.9 (400/1434) | 38.4 (551/1434) |
| P Value | Reference | <.001 | <.001 | <.001 | .10 | <.001 |
| Clinically Acceptable Reports by Standard Criterion, % | 74.3 (1065/1434) | 84.5 (1212/1434) | 41.1 (589/1434) | 65.6 (941/1434) | 71.4 (1024/1434) | 63.7 (914/1434) |
| P Value | Reference | <.001 | <.001 | <.001 | .04 | <.001 |
| Clinically Acceptable Reports by Stringent Criterion, % | 43.5 (624/1434) | 58.2 (834/1434) | 25.3 (363/1434) | 37.3 (535/1434) | 42.9 (615/1434) | 39.7 (570/1434) |
| P Value | Reference | <.001 | <.001 | <.001 | .68 | .01 |
| Hallucination, %* | 0.1 (1/1425) | 0.3 (4/1425) | 11.0 (157/1425) | 17.4 (248/1425) | 5.4 (77/1425) | 12.3 (175/1425) |
| P Value | Reference | .21 | <.001 | <.001 | <.001 | <.001 |
| Clarity of Language, % | 78.1 (1120/1434) | 82.9 (1189/1434) | 88.0 (1262/1434) | 78.4 (1124/1434) | 69.7 (1000/1434) | 88.4 (1268/1434) |
| P Value | Reference | .001 | <.001 | .85 | <.001 | <.001 |

Clinical acceptability was defined as “acceptable as is” or “with minor revision” (standard), and strictly as “acceptable as is” only (stringent).

Clarity of language was defined as a score of 4 or higher (good to excellent).

P values represent comparisons against the radiologist as the reference.

*Three patients whose original reports contained measurements were excluded from the analysis of hallucinations.

**Table 3:** Diagnostic Performance with Chest CT as the Reference Standard

| | Radiologist | | AIRead | | Lingshu | |
|---|---|---|---|---|---|---|
| **Findings** | **Sensitivity, %** | **Specificity, %** | **Sensitivity, %** | **Specificity, %** | **Sensitivity, %** | **Specificity, %** |
| Aortic aneurysm | 0 (0, 41.0) [0/7] | 100 (99.2, 100) [471/471] | 28.6 (3.7, 71.0) [2/7] | 98.1 (96.3, 99.1) [462/471] | 0 (0, 41.0) [0/7] | 100 (99.2, 100) [471/471] |
| Cardiomegaly | 21.3 (13.0, 32.6) [16/75] | 98.3 (96.3, 99.2) [396/403] | 86.7 (76.4, 93.1) [65/75] | 77.9 (73.5, 81.8) [314/403] | 86.7 (76.4, 93.1) [65/75] | 61.5 (56.6, 66.3) [248/403] |
| Cavity | 20 (4.3, 48.1) [3/15] | 100 (99.2, 100) [463/463] | 6.7 (0.2, 31.9) [1/15] | 100 (99.2, 100) [463/463] | 0 (0, 21.8) [0/15] | 100 (99.2, 100) [463/463] |
| Emphysema | 6.0 (2.2, 14.0) [5/84] | 100 (99.1, 100) [394/394] | 15.5 (8.8, 25.4) [13/84] | 99.2 (97.8, 99.8) [391/394] | 2.4 (0.3, 8.3) [2/84] | 95.7 (93.0, 97.4) [377/394] |
| Hilar enlargement | 30 (6.7, 65.2) [3/10] | 99.1 (97.8, 99.8) [464/468] | 80 (44.4, 97.5) [8/10] | 94.7 (92.1, 96.4) [443/468] | 0 (0, 30.8) [0/10] | 98.9 (97.4, 99.6) [463/468] |
| Lung opacity (consolidation or GGO) | 60.7 (54.4, 66.7) [156/257] | 85.1 (79.5, 89.4) [188/221] | 77.8 (72.1, 82.6) [200/257] | 70.6 (64.0, 76.4) [156/221] | 16.3 (12.2, 21.6) [42/257] | 89.1 (84.1, 92.8) [197/221] |
| Mediastinal widening | 21.1 (6.1, 45.6) [4/19] | 99.3 (98.1, 99.9) [456/459] | 57.9 (33.5, 79.7) [11/19] | 99.1 (97.8, 99.8) [455/459] | 15.8 (3.4, 39.6) [3/19] | 85.4 (81.8, 88.4) [392/459] |
| Miliary nodule | 0 (0, 30.8) [0/10] | 100 (99.2, 100) [468/468] | 0 (0, 30.8) [0/10] | 100 (99.2, 100) [468/468] | 0 (0, 30.8) [0/10] | 100 (99.2, 100) [468/468] |
| Nodule or mass | 20.5 (13.7, 29.4) [23/112] | 97.0 (94.5, 98.4) [355/366] | 36.6 (27.9, 46.3) [41/112] | 87.4 (83.5, 90.6) [320/366] | 0.9 (0, 4.9) [1/112] | 99.5 (98.0, 99.9) [364/366] |
| Pericardial effusion | 0 (0, 6.1) [0/59] | 100 (99.1, 100) [419/419] | 0 (0, 6.1) [0/59] | 100 (99.1, 100) [419/419] | 0 (0, 6.1) [0/59] | 100 (99.1, 100) [419/419] |
| Pleural effusion | 43.1 (36.2, 50.2) [87/202] | 97.5 (94.6, 98.9) [269/276] | 68.3 (61.4, 74.6) [138/202] | 92.0 (88.0, 94.8) [254/276] | 21.3 (16.0, 27.7) [43/202] | 90.2 (85.9, 93.3) [249/276] |
| Pneumomediastinum | 100 (2.5, 100) [1/1] | 100 (99.2, 100) [477/477] | 100 (2.5, 100) [1/1] | 100 (99.2, 100) [477/477] | 0 (0, 97.5) [0/1] | 100 (99.2, 100) [477/477] |
| Pneumoperitoneum | 0 (0, 97.5) [0/1] | 100 (99.2, 100) [477/477] | 0 (0, 97.5) [0/1] | 99.8 (98.8, 100) [476/477] | 0 (0, 97.5) [0/1] | 100 (99.2, 100) [477/477] |
| Pneumothorax | 77.8 (40.0, 97.2) [7/9] | 99.8 (98.8, 100) [468/469] | 77.8 (40.0, 97.2) [7/9] | 100 (99.2, 100) [469/469] | 0 (0, 33.6) [0/9] | 99.6 (98.5, 99.9) [467/469] |
| Reticular opacity | 7.1 (0.9, 23.5) [2/28] | 98.4 (96.7, 99.3) [443/450] | 25.0 (10.7, 44.9) [7/28] | 99.8 (98.8, 100) [449/450] | 3.6 (0.1, 18.3) [1/28] | 95.6 (93.1, 97.2) [430/450] |

Values in parentheses indicate 95% confidence intervals, and data in brackets denote numerators and denominators.

GGO = ground-glass opacity.

**Table 4:** Diagnostic Performance with Chest CT as the Reference Standard

| | MAIRA-2 | | MedGemma | | MedVersa | |
|---|---|---|---|---|---|---|
| **Findings** | **Sensitivity,** | **Specificity,** | **Sensitivity,** | **Specificity,** | **Sensitivity,** | **Specificity,** |
| Aortic aneurysm | 0 (0, 41.0) [0/7] | 100 (99.2, 100) [471/471] | 0 (0, 41.0) [0/7] | 100 (99.2, 100) [471/471] | 0 (0, 41.0) [0/7] | 100 (99.2, 100) [471/471] |
| Cardiomegaly | 72.0 (60.3, 81.5) [54/75] | 85.6 (81.7, 88.8) [345/403] | 73.3 (61.7, 82.6) [55/75] | 81.1 (76.9, 84.8) [327/403] | 69.3 (57.5, 79.2) [52/75] | 87.3 (83.6, 90.4) [352/403] |
| Cavity | 0 (0, 21.8) [0/15] | 99.8 (98.8, 100) [462/463] | 0 (0, 21.8) [0/15] | 99.6 (98.4, 99.9) [461/463] | 0 (0, 21.8) [0/15] | 100 (99.2, 100) [463/463] |
| Emphysema | 6.0 (2.2, 14.0) [5/84] | 98.2 (96.2, 99.2) [387/394] | 4.8 (1.3, 11.7) [4/84] | 97.5 (95.2, 98.7) [384/394] | 20.2 (12.6, 30.7) [17/84] | 94.7 (91.8, 96.6) [373/394] |
| Hilar enlargement | 20 (2.5, 55.6) [2/10] | 96.8 (94.6, 98.1) [453/468] | 10 (0.3, 44.5) [1/10] | 96.2 (93.9, 97.6) [450/468] | 10 (0.3, 44.5) [1/10] | 98.3 (96.5, 99.2) [460/468] |
| Lung opacity (consolidation or GGO) | 51.4 (45.1, 57.6) [132/257] | 82.8 (77.0, 87.4) [183/221] | 76.7 (70.9, 81.6) [197/257] | 47.5 (40.8, 54.3) [105/221] | 39.3 (33.3, 45.6) [101/257] | 86.9 (81.5, 90.9) [192/221] |
| Mediastinal widening | 5.3 (0.1, 26.0) [1/19] | 99.3 (98.1, 99.9) [456/459] | 15.8 (3.4, 39.6) [3/19] | 95.6 (93.2, 97.2) [439/459] | 10.5 (1.3, 33.1) [2/19] | 99.6 (98.4, 99.9) [457/459] |
| Miliary nodule | 0 (0, 30.8) [0/10] | 100 (99.2, 100) [468/468] | 0 (0, 30.8) [0/10] | 100 (99.2, 100) [468/468] | 0 (0, 30.8) [0/10] | 100 (99.2, 100) [468/468] |
| Nodule or mass | 7.1 (3.4, 14.0) [8/112] | 98.1 (95.9, 99.2) [359/366] | 26.8 (19.1, 36.1) [30/112] | 82.8 (78.4, 86.4) [303/366] | 3.6 (1.0, 8.9) [4/112] | 98.4 (96.3, 99.3) [360/366] |
| Pericardial effusion | 0 (0, 6.1) [0/59] | 100 (99.1, 100) [419/419] | 5.1 (1.1, 14.1) [3/59] | 98.1 (96.1, 99.1) [411/419] | 0 (0, 6.1) [0/59] | 100 (99.1, 100) [419/419] |
| Pleural effusion | 43.6 (36.7, 50.7) [88/202] | 91.3 (87.2, 94.2) [252/276] | 71.8 (65.0, 77.8) [145/202] | 75.0 (69.4, 79.9) [207/276] | 48.0 (41.0, 55.1) [97/202] | 97.1 (94.2, 98.6) [268/276] |
| Pneumomediastinum | 0 (0, 97.5) [0/1] | 100 (99.2, 100) [477/477] | 0 (0, 97.5) [0/1] | 100 (99.2, 100) [477/477] | 0 (0, 97.5) [0/1] | 100 (99.2, 100) [477/477] |
| Pneumoperitoneum | 0 (0, 97.5) [0/1] | 100 (99.2, 100) [477/477] | 0 (0, 97.5) [0/1] | 100 (99.2, 100) [477/477] | 0 (0, 97.5) [0/1] | 100 (99.2, 100) [477/477] |
| Pneumothorax | 55.6 (21.2, 86.3) [5/9] | 99.6 (98.5, 99.9) [467/469] | 22.2 (2.8, 60.0) [2/9] | 98.1 (96.3, 99.1) [460/469] | 44.4 (13.7, 78.8) [4/9] | 99.4 (98.1, 99.9) [466/469] |
| Reticular opacity | 7.1 (0.9, 23.5) [2/28] | 99.1 (97.7, 99.8) [446/450] | 28.6 (13.2, 48.7) [8/28] | 93.1 (90.3, 95.2) [419/450] | 28.6 (13.2, 48.7) [8/28] | 95.6 (93.1, 97.2) [430/450] |

Values in parentheses indicate 95% confidence intervals, and data in brackets denote numerators and denominators.

GGO = ground-glass opacity.

## Figure Legends

**Figure 1:** Flow diagram of patient inclusion and exclusion.

**Figure 2:** Comparative evaluation of chest radiograph reports generated by vision–language models against real-world radiologist-written reports.

All chest radiograph reports were independently reviewed and rated in a blinded and randomized manner by three thoracic radiologists.

**(A)** Distribution of RADPEER scores across five vision–language models (AIRead, Lingshu, MAIRA-2, MedGemma, and MedVersa) and radiologists (i.e., original reports). RADPEER categories indicate levels of interpretive agreement, with lower scores reflecting higher agreement and higher scores reflecting greater discrepancies. Within each score, the "a" descriptor denotes a clinically insignificant discrepancy, whereas "b" denotes a clinically significant one.

**(B)** Clinical acceptability of reports, categorized as acceptable as is, acceptable with minor revision, acceptable only with major revision, or not acceptable.

**(C)** Clarity of language scored on a 5-point scale with higher scores representing greater clarity.

**Figure 3:** Representative chest radiographs and reports of true-positive cases.

**(A)** A 74-year-old woman with dyspnea. Frontal chest radiograph shows multifocal patchy ground-glass opacities and consolidation, predominantly in the right lung, with bilateral costophrenic angle blunting suggestive of bilateral pleural effusion. CT confirmed corresponding abnormalities. All report sources appropriately detected lung opacity (with variable terminology); however, bilateral pleural effusion was missed in some reports (radiologist, Lingshu, MedVersa). Green=true positive/true negative; red=false positive/false negative.

**(B)** A 79-year-old man with dyspnea. Frontal chest radiograph shows increased pulmonary vascularity with equivocal peribronchial opacities; CT confirmed interlobular septal thickening. Some report sources suggested pulmonary congestion/edema, whereas MAIRA-2 and MedVersa failed to report these findings. Green=true positive/true negative; red=false positive/false negative.

**Figure 4:** Representative chest radiographs and reports of false-positive cases.

**(A)** A 35-year-old man with chest discomfort. Radiograph demonstrates left pleural effusion and cardiomegaly; CT confirmed pericardial effusion with pericardial thickening and left pleural effusion. AIRead and MedVersa correctly described these findings, whereas others partially or entirely missed them. MedVersa falsely reported a pneumothorax, and Lingshu generated comparison statements. Green=true positive/true negative; red=false positive/false negative; yellow=hallucination.

**(B)** A 66-year-old woman with fever. Chest radiograph was interpreted as normal by all readers; CT revealed subtle bilateral ground-glass opacities, suggestive of atypical pneumonia. Lingshu reported endotracheal intubation, which was a false positive. Green=true positive/true negative; red=false positive/false negative; yellow=hallucination.

**Figure 5:** Representative chest radiographs and reports illustrating hallucinations and issues of language clarity.

**(A)** A 29-year-old man with chest discomfort. Radiograph and CT showed pneumomediastinum and subcutaneous emphysema. Radiologist and AIRead reports were appropriate, whereas MAIRA-2 included comparison statements and MedGemma inferred medical information. Green=true positive/true negative; red=false positive/false negative; yellow=hallucination.

**(B)** A 26-year-old man with chest pain and pneumothorax. Lingshu failed to report the large right pneumothorax, while other models described it correctly. MedVersa included a hallucinated statement regarding physician communication. Notably, MedGemma employed an enumerative manner, incorrectly characterizing the obvious compressive atelectasis caused by a large pneumothorax as mass/tumor, infection, fluid, or consolidation, thereby obscuring language clarity. Additionally, MedGemma's reports sometimes contained incomplete sentences. Green=true positive/true negative; red=false positive/false negative; yellow=hallucination/poor language clarity; underline=incomplete sentence.

**Figure 6:** Sensitivity and specificity for individual findings.

Forest plots show the sensitivity (left) and specificity (right) with 95% confidence intervals for 15 findings, including cardiomegaly, consolidation or ground-glass opacity (GGO), pleural effusion, pneumothorax, pneumomediastinum, hilar enlargement, mediastinal widening, reticular opacity, nodule or mass, emphysema, aortic aneurysm, cavity, pericardial effusion, miliary nodule, and pneumoperitoneum. Results are compared across five vision–language models (AIRead, Lingshu, MAIRA-2, MedGemma, and MedVersa) and radiologists (i.e., original reports).

# Figures

**Figure 1:** Flow diagram of patient inclusion and exclusion.

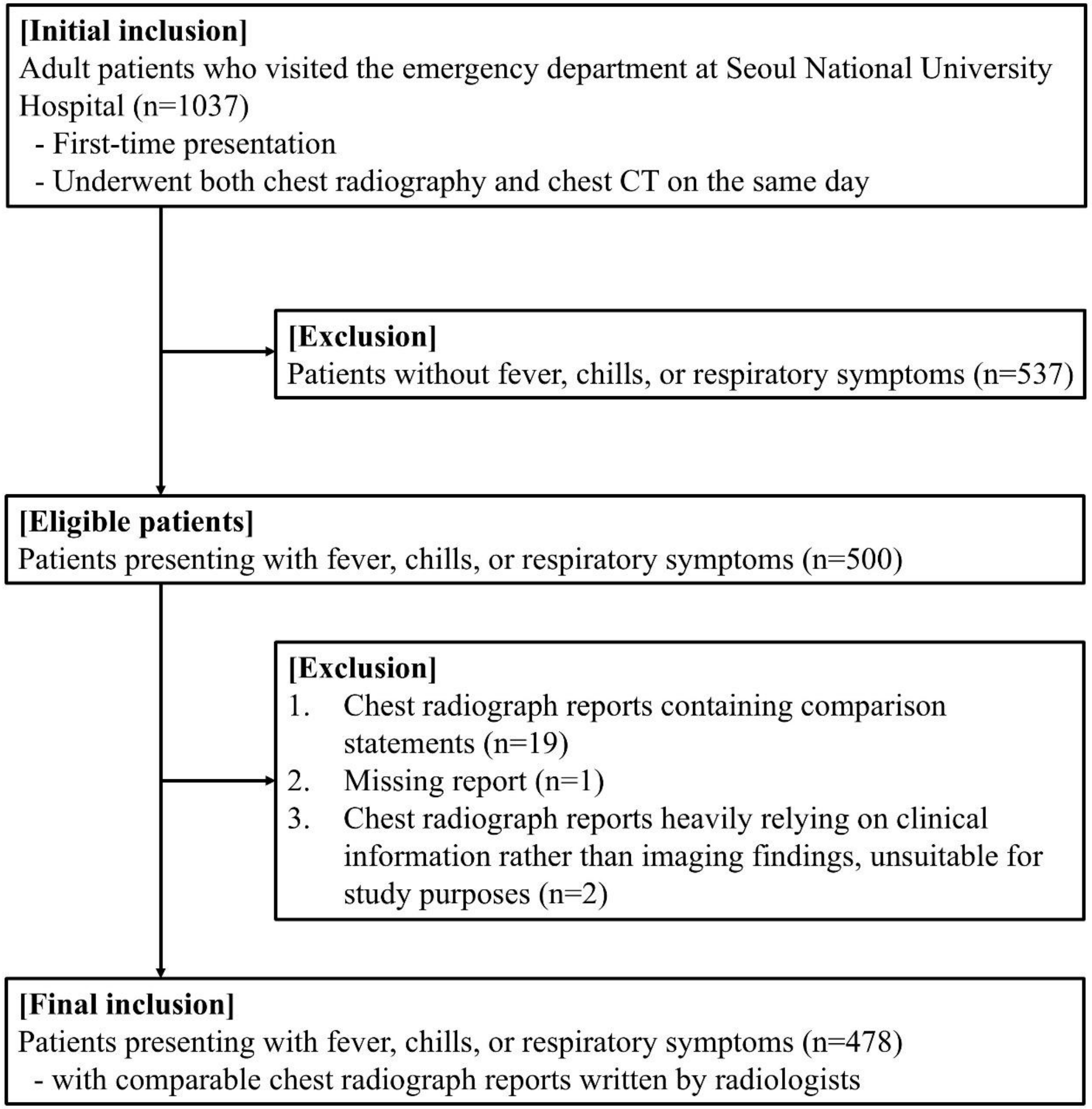

**Figure 2:** Comparative evaluation of chest radiograph reports generated by vision–language models against real-world radiologist-written reports.

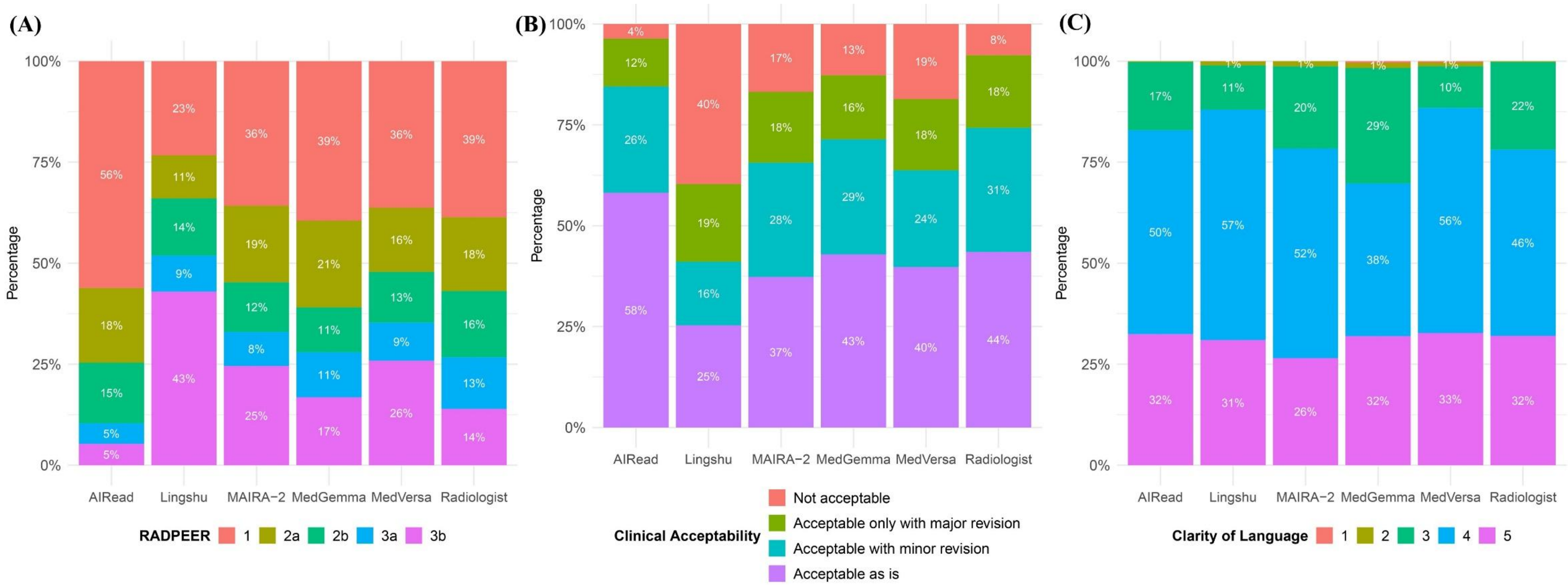

All chest radiograph reports were independently reviewed and rated in a blinded and randomized manner by three thoracic radiologists.

**(A)** Distribution of RADPEER scores across five vision–language models (AIRead, Lingshu, MAIRA-2, MedGemma, and MedVersa) and radiologists (i.e., original reports). RADPEER categories indicate levels of interpretive agreement, with lower scores reflecting higher agreement and higher scores reflecting greater discrepancies. Within each score, the "a" descriptor denotes a clinically insignificant discrepancy, whereas "b" denotes a clinically significant one.

**(B)** Clinical acceptability of reports, categorized as acceptable as is, acceptable with minor revision, acceptable only with major revision, or not acceptable.

**(C)** Clarity of language scored on a 5-point scale with higher scores representing greater clarity.

**Figure 3:** Representative chest radiographs and reports of true-positive cases.

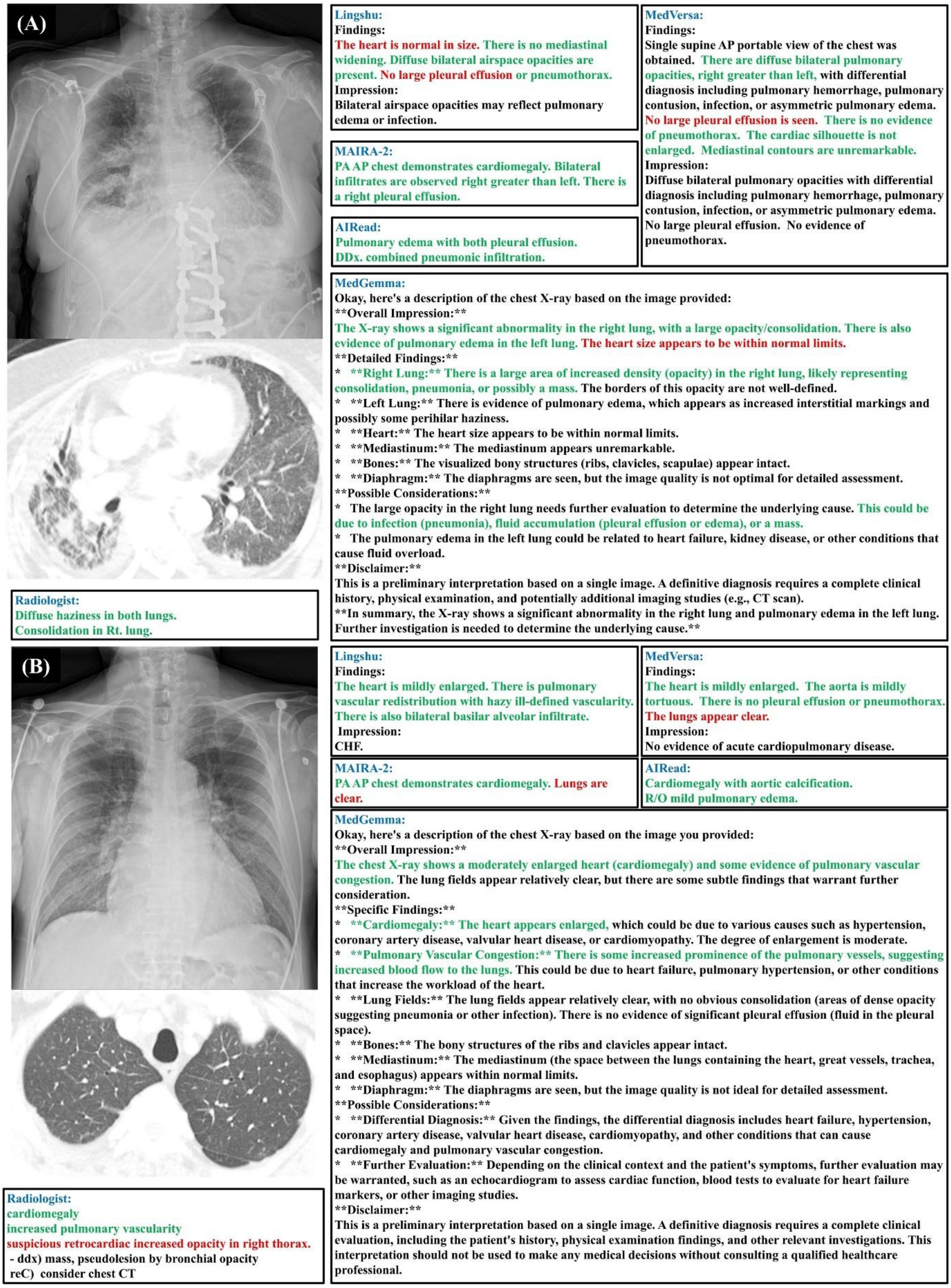

**(A)** A 74-year-old woman with dyspnea. Frontal chest radiograph shows multifocal patchy ground-glass opacities and consolidation, predominantly in the right lung, with bilateral costophrenic angle blunting suggestive of bilateral pleural effusion. CT confirmed corresponding abnormalities. All report sources appropriately detected lung opacity (with variable terminology); however, bilateral pleural effusion was missed in some reports (radiologist, Lingshu, MedVersa). Green=true positive/true negative; red=false positive/false negative.

**(B)** A 79-year-old man with dyspnea. Frontal chest radiograph shows increased pulmonary vascularity with equivocal peribronchial opacities; CT confirmed interlobular septal thickening. Some report sources suggested pulmonary congestion/edema, whereas MAIRA-2 and MedVersa failed to report these findings. Green=true positive/true negative; red=false positive/false negative.

**Figure 4:** Representative chest radiographs and reports of false-positive cases.

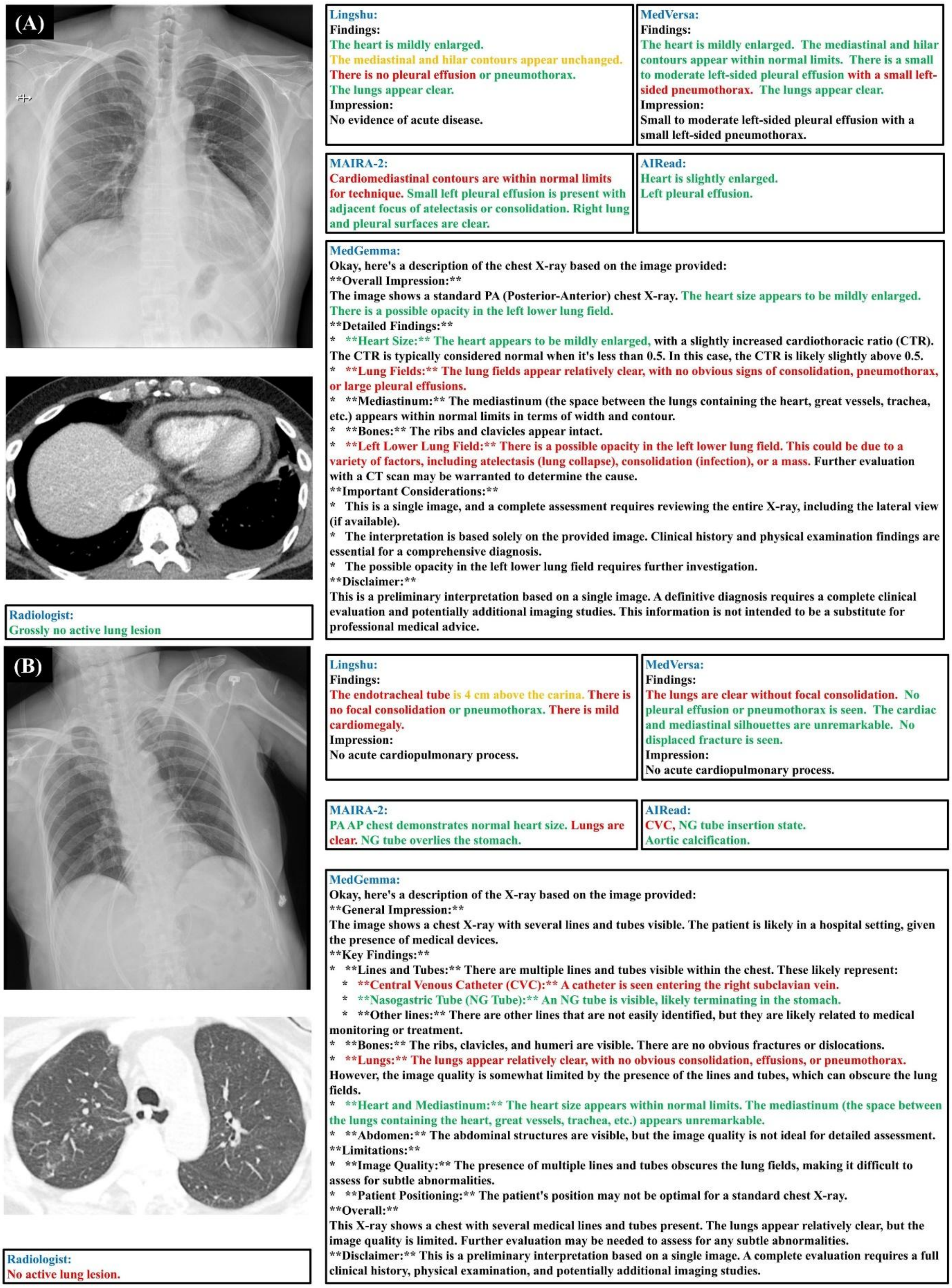

**(A)** A 35-year-old man with chest discomfort. Radiograph demonstrates left pleural effusion and cardiomegaly; CT confirmed pericardial effusion with pericardial thickening and left pleural effusion. AIRead and MedVersa correctly described these findings, whereas others partially or entirely missed them. MedVersa falsely reported a pneumothorax, and Lingshu generated comparison statements. Green=true positive/true negative; red=false positive/false negative; yellow=hallucination.

**(B)** A 66-year-old woman with fever. Chest radiograph was interpreted as normal by all readers; CT revealed subtle bilateral ground-glass opacities, suggestive of atypical pneumonia. Lingshu reported endotracheal intubation, which was a false positive. Green=true positive/true negative; red=false positive/false negative; yellow=hallucination.

**Figure 5:** Representative chest radiographs and reports illustrating hallucinations and issues of language clarity.

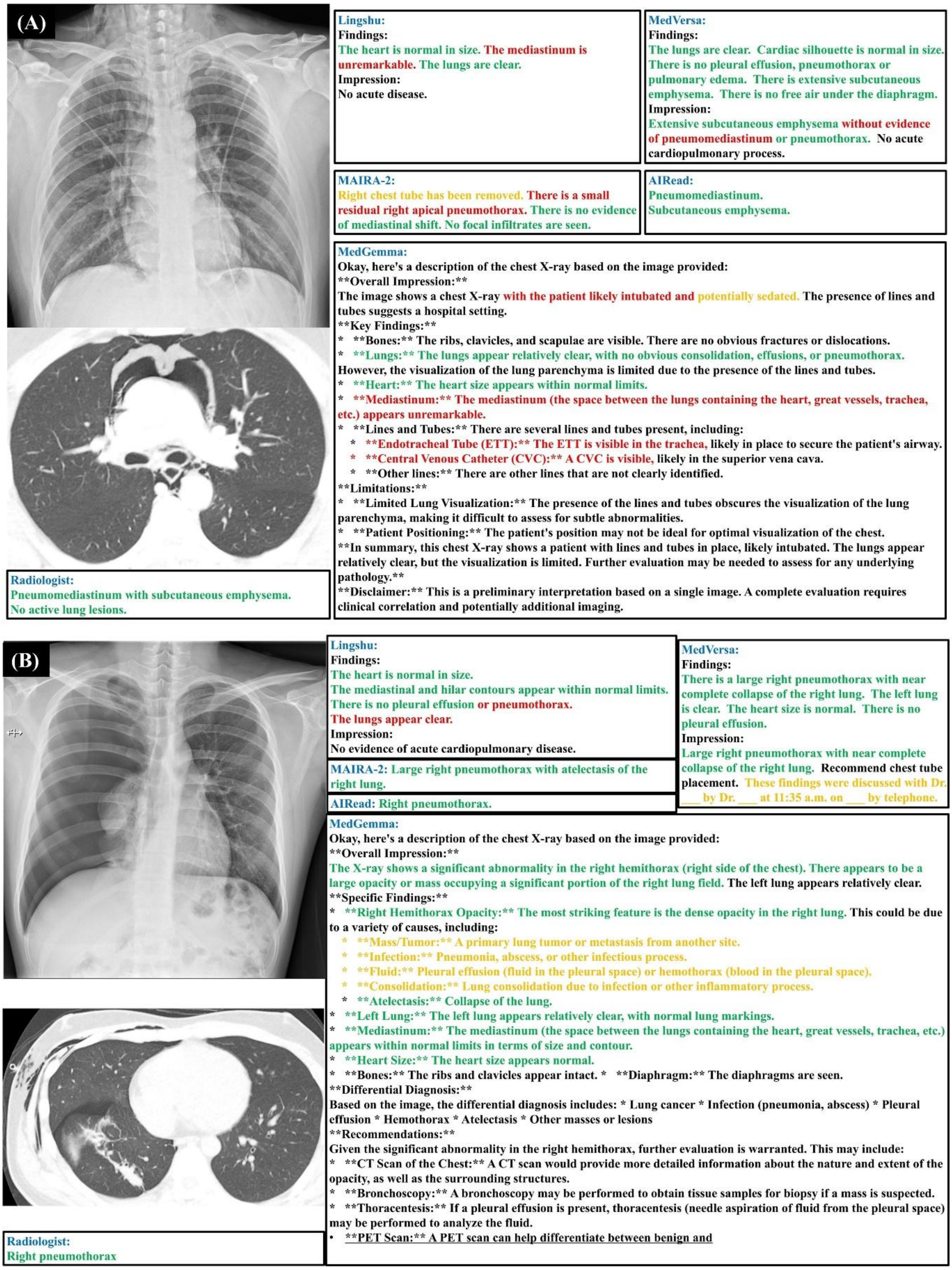

**(A)** A 29-year-old man with chest discomfort. Radiograph and CT showed pneumomediastinum and subcutaneous emphysema. Radiologist and AIRead reports were appropriate, whereas MAIRA-2 included comparison statements and MedGemma inferred medical information. Green=true positive/true negative; red=false positive/false negative; yellow=hallucination.

**(B)** A 26-year-old man with chest pain and pneumothorax. Lingshu failed to report the large right pneumothorax, while other models described it correctly. MedVersa included a hallucinated statement regarding physician communication. Notably, MedGemma employed an enumerative manner, incorrectly characterizing the obvious compressive atelectasis caused by a large pneumothorax as mass/tumor, infection, fluid, or consolidation, thereby obscuring language clarity. Additionally, MedGemma's reports sometimes contained incomplete sentences. Green=true positive/true negative; red=false positive/false negative; yellow=hallucination/poor language clarity; underline=incomplete sentence.

**Figure 6:** Sensitivity and specificity for individual findings.

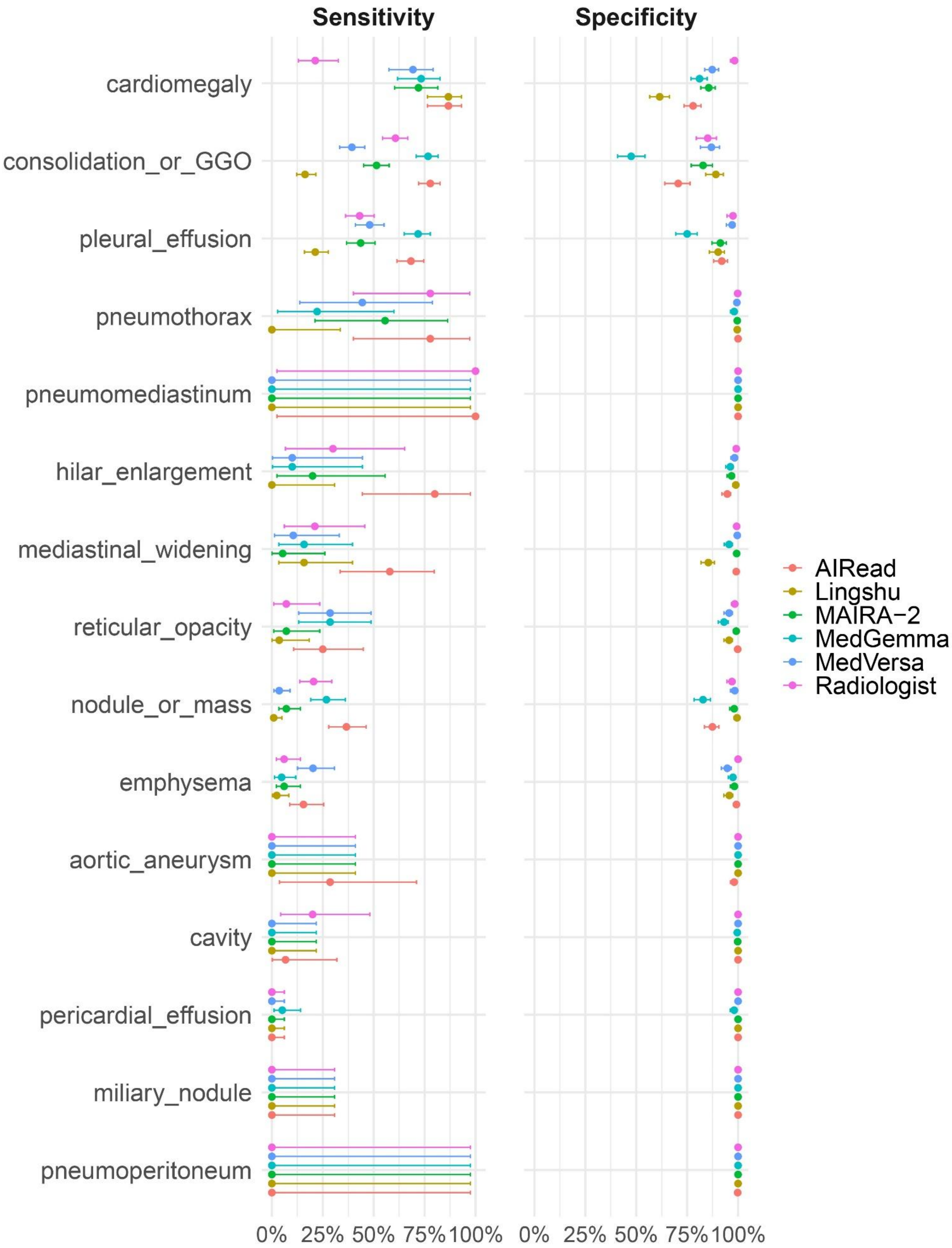

Forest plots show the sensitivity (left) and specificity (right) with 95% confidence intervals for 15 findings, including cardiomegaly, consolidation or ground-glass opacity (GGO), pleural effusion, pneumothorax, pneumomediastinum, hilar enlargement, mediastinal widening, reticular opacity, nodule or mass, emphysema, aortic aneurysm, cavity, pericardial effusion, miliary nodule, and pneumoperitoneum. Results are compared across five vision–language models (AIRead, Lingshu, MAIRA-2, MedGemma, and MedVersa) and radiologists (i.e., original reports).

# Supplementary Texts

### Supplementary Text 1: Medical image–specific vision–language models

**1) AIRead:** AIRead was designed to generate reports from frontal chest radiographs (CXRs) (either anteroposterior or posteroanterior view). The model comprises an image encoder and a language model; the encoder localizes potential abnormal regions and assigns 61 finding categories, and these outputs are passed to the language model for report generation. The model contains 2.6 billion parameters. The performance of an early version has been reported previously[1,2]. The initial training dataset comprised approximately 14 million CXRs paired with reports, retrospectively collected from 11 tertiary hospitals in Korea and from multiple hospitals and clinics in the United States, of which 8 million image–report pairs were used to train the image encoder. Unlike the other models used in our experiments, the parameters of AIRead are not publicly available. However, the model can be accessed at https://airead.soombit.ai/, and the version used in this study was v0.2.5. This model is for research use only and has not been cleared or approved by regulatory authorities for clinical use.

**2) Lingshu:** Lingshu emphasizes comprehensive data curation to enhance medical artificial intelligence reasoning across multiple tasks, including multimodal question answering, text-based question answering, and medical report generation[3]. The model accepts diverse imaging modalities such as plain radiograph, CT, MRI, ultrasound, dermoscopy, fundus photography, histopathology, and microscopy. The training dataset comprised 2.55 million instances, with x-ray images accounting for 13%. Lingshu was built on the Qwen2.5-VL-Instruct model, with 7 billion and 32 billion parameters. In a pilot evaluation with a thoracic radiologist, the 32B model performed comparably to the 7B model for CXR report generation. Therefore, for these experiments, we used the 7B parameters available at https://huggingface.co/lingshu-medical-mllm/Lingshu-7B, and followed the inference recommendations provided in the model card.

**3) MAIRA-2:** MAIRA-2 was developed to provide comprehensive support for radiologists in interpreting CXRs by leveraging both frontal and lateral views, prior studies, and clinical indication[4]. The model further enhances report interpretation through a novel task termed grounded report generation, which explicitly localizes each finding within the image. MAIRA-2 consists of an image encoder and a large language model with a visual adapter, using the radiology-specialized Rad-DINO-MAIRA-2 as the image encoder and Vicuna 7B v1.5 as the language model. The total model size is approximately 6.9 billion parameters. The training dataset included 510K image–report pairs from MIMIC-CXR, PadChest, USMix, and IU-Xray. For these experiments, we used the model parameters available at https://huggingface.co/microsoft/maira-2. For inference, we used "without grounding option", since the quality of the report only differs slightly as described in the model card.

**4) MedGemma:** MedGemma is a medical foundation model designed to perform across diverse clinical tasks while reducing task-specific tuning requirements, thereby accelerating the development of artificial intelligence for healthcare applications[5]. Its vision module processes four imaging domains: radiology (CXRs, CT, MRI), histopathology, dermatology, and ophthalmology. Supported tasks include medical visual question answering, medical image classification, CXR report generation, and related applications. The architecture comprises a visual encoder and a language model connected via a visual adapter; the visual encoder is fine-tuned from SigLIP, and the language model is based on Gemma 3. Similar to MedVersa, MIMIC-CXR was the primary training dataset for CXR report generation. MedGemma is released in multiple variants, including instruction-tuned and pre-trained models, with parameter sizes ranging from 4 billion to 27 billion. At the start of this study, the 27B multimodal model was not available; therefore, we used the 4B instruction-tuned model (https://huggingface.co/google/medgemma-4b-it) and followed the inference recommendations provided in the model card.

**5) MedVersa:** MedVersa is a medical artificial intelligence generalist developed for diverse tasks in medical image interpretation[6]. It was trained on 13 million annotated instances spanning 11 tasks across 3 imaging modalities: CXRs, CT scans, and dermoscopic images. Supported tasks include report generation, open-ended visual question answering, and related applications. For CXR report generation, the primary training dataset was MIMIC-CXR. After filtering, MedVersa utilized approximately 122K CXRs paired with reports. The model consists of a 2D/3D visual encoder and a large language model, with an adapter employed to transfer visual information to the language model. The total model size is approximately 6.9 billion parameters. For these experiments, we used the model parameters available at https://huggingface.co/hyzhou/MedVersa_Internal. For CXR report generation, we followed the recommendations provided in the model card, including the specified environment setup.

**6) Inference settings:** All inferences were performed on a single NVIDIA H100 graphics processing unit with CUDA 12.4. To minimize the effects of stochastic variation in the generation process, greedy decoding was used to generate reports across all methods. Because the evaluation of this study focused on reports from a single frontal CXR, models capable of incorporating prior reports, additional images (including lateral views), or clinical information (e.g., age) were instead provided with blank or no inputs for these fields.

**Supplementary Text 2: Details of a large language model–based labeler for extracting abnormal findings from free-text chest radiograph reports**

A practical approach to quantitatively assess report-generation models is to compare labels extracted from model-generated and radiologist-written free-text reports. To this end, model-based labelers such as RadGraph and CheXpert++ have been developed to extract predefined labels and relations from radiology reports[7,8]. More recently, leveraging the medical knowledge embedded in large language models (LLMs) has emerged as a promising direction for label extraction. In this line of work, CheXGPT utilized LLM-based medical knowledge to extract 13 labels (atelectasis, consolidation, effusion, fracture, hyperinflation, lung opacity, nodule, pleural lesion, pneumothorax, pulmonary edema, subcutaneous emphysema, subdiaphragmatic gas, widened mediastinal silhouette)[9]. CheXGPT also distilled the knowledge of a LLM into a lightweight language model, thereby enabling efficient label extraction and supporting a wide range of clinical applications.

For this study, we developed a labeler similar to CheXGPT by leveraging LLMs to extract labels, but the coverage extended from 13 to 76 abnormal categories. We adopted a similar training strategy, in which pseudo-label sets were first extracted from chest radiograph reports by using a LLM. These pseudo-labels were refined through domain-specific post-processing and subsequently used to supervise the training of a Bidirectional Encoder Representations from Transformers-based extraction model[10].

# Supplementary Tables

**Supplementary Table 1:** Summarization of Evaluation Metrics

| **Criteria** | **Details** |
|---|---|
| **RADPEER score** | |
| **1** | Agree with interpretations. |
| **2** | Disagree with interpretations which are understandable misses. |
| | **a:** unlikely to be clinically significant. |
| | **b**: likely to be clinically significant. |
| **3** | Disagree with interpretations which are unacceptable misses. |
| | **a:** unlikely to be clinically significant. |
| | **b:** likely to be clinically significant. |
| **Clinical acceptability** | |
| **1: Not acceptable** | The report contains major factual errors (e.g., missed referable findings, false-positive findings, incorrect laterality or diagnosis) that make it unsafe for clinical use. The content is misleading or unreliable and cannot be used even after editing. |
| **2: Acceptable only with major revision** | The report contains one or more significant problems (e.g., omission of a finding, vague or clinically misleading wording, or potentially harmful misinterpretation) that require extensive edits across multiple sentences or restructuring. A radiologist would need to spend meaningful time reviewing and rewriting the report before it could be used. |
| **3: Acceptable only with minor revision** | The report is generally accurate and includes key findings, but may have minor issues such as slightly awkward phrasing, limited detail (e.g., lacking severity), or subtle clarification needs. It can be used after quick edits without changing core content. |
| **4: Acceptable as is** | The report is accurate, clear, and clinically actionable as written. It requires no edits and could be signed and submitted directly. |
| **Hallucination** | The presence of information that could not be derived solely from a single frontal radiograph, in which |
| | information on scale was not available (e.g., references to a lateral view, clinical history, comparative statements, or size or distance descriptions). |
| **Language clarity** | Subjective scoring using a 5-point scale (1 = unacceptable, 2 = poor, 3 = moderate, 4 = good, 5 = excellent). |

**Supplementary Table 2:** Subgroup Analysis in Men

| Variable | Radiologist | AIRead | Lingshu | MAIRA-2 | MedGemma | MedVersa |
|---|---|---|---|---|---|---|
| RADPEER Score 3b, % | 13.4 (113/846) | 5.6 (47/846) | 42.9 (363/846) | 24.2 (205/846) | 17.1 (145/846) | 24.7 (209/846) |
| P Value | Reference | <.001 | <.001 | <.001 | .01 | <.001 |
| RADPEER Score 2b or 3b, % | 27.8 (235/846) | 18.8 (159/846) | 55.6 (470/846) | 34.5 (292/846) | 27.5 (233/846) | 35.2 (298/846) |
| P Value | Reference | <.001 | <.001 | <.001 | .90 | <.001 |
| Clinically Acceptable Reports by Standard Criterion, % | 76.4 (646/846) | 85.2 (721/846) | 42.3 (358/846) | 67.5 (571/846) | 71.7 (607/846) | 67.1 (568/846) |
| P Value | Reference | <.001 | <.001 | <.001 | .01 | <.001 |
| Clinically Acceptable Reports by Stringent Criterion, % | 46.2 (391/846) | 59.6 (504/846) | 26.7 (226/846) | 38.1 (322/846) | 42.7 (361/846) | 42.7 (361/846) |
| P Value | Reference | <.001 | <.001 | <.001 | .07 | .07 |
| Hallucination, % | 0.1 (1/846) | 0.4 (3/846) | 6.9 (58/846) | 19.3 (163/846) | 5.0 (42/846) | 11.5 (97/846) |
| P Value | Reference | .33 | <.001 | <.001 | <.001 | <.001 |
| Clarity of Language, % | 76.6 (648/846) | 82.9 (701/846) | 89.2 (755/846) | 77.9 (659/846) | 69.4 (587/846) | 87.9 (744/846) |
| P Value | Reference | .001 | <.001 | .52 | .001 | <.001 |

Clinical acceptability was defined as “acceptable as is” or “with minor revision” (standard), and strictly as “acceptable as is” only (stringent).

Clarity of language was defined as a score of 4 or higher (good to excellent).

P values represent comparisons against the radiologist as the reference.

**Supplementary Table 3:** Subgroup Analysis in Women

| Variable | Radiologist | AIRead | Lingshu | MAIRA-2 | MedGemma | MedVersa |
|---|---|---|---|---|---|---|
| RADPEER Score 3b, % | 14.8 (87/588) | 4.9 (29/588) | 43.0 (253/588) | 25.0 (147/588) | 16.3 (96/588) | 27.6 (162/588) |
| P Value | Reference | <.001 | <.001 | <.001 | .41 | <.001 |
| RADPEER Score 2b or 3b, % | 34.0 (200/588) | 22.4 (132/588) | 59.4 (349/588) | 40.1 (236/588) | 28.4 (167/588) | 43.0 (253/588) |
| P Value | Reference | <.001 | <.001 | .01 | .02 | <.001 |
| Clinically Acceptable Reports by Standard Criterion, % | 71.3 (419/588) | 83.5 (491/588) | 39.3 (231/588) | 62.9 (370/588) | 70.9 (417/588) | 58.8 (346/588) |
| P Value | Reference | <.001 | <.001 | <.001 | .88 | <.001 |
| Clinically Acceptable Reports by Stringent Criterion, % | 39.6 (233/588) | 56.1 (330/588) | 23.3 (137/588) | 36.2 (213/588) | 43.2 (254/588) | 35.5 (209/588) |
| P Value | Reference | <.001 | <.001 | .15 | .14 | .08 |
| Hallucination, %* | 0 (0/579) | 0.2 (1/579) | 17.1 (99/579) | 14.7 (85/579) | 6.0 (35/579) | 13.5 (78/579) |
| P Value | Reference | N/A | N/A | N/A | N/A | N/A |
| Clarity of Language, % | 80.3 (472/588) | 83.0 (488/588) | 86.2 (507/588) | 79.1 (465/588) | 70.2 (413/588) | 89.1 (524/588) |
| P Value | Reference | .22 | .006 | .61 | <.001 | <.001 |

Clinical acceptability was defined as “acceptable as is” or “with minor revision” (standard), and strictly as “acceptable as is” only (stringent).

Clarity of language was defined as a score of 4 or higher (good to excellent).

P values represent comparisons against the radiologist as the reference.

*Three patients whose original reports contained measurements were excluded from the analysis of hallucinations.

N/A = not available.

**Supplementary Table 4:** Subgroup Analysis of Patients Aged 65 Years and Older

| **Variable** | **Radiologist** | **AIRead** | **Lingshu** | **MAIRA-2** | **MedGemma** | **MedVersa** |
|---|---|---|---|---|---|---|
| RADPEER Score 3b, % | 17.7 (140/792) | 5.7 (45/792) | 47.5 (376/792) | 30.7 (243/792) | 19.2 (152/792) | 29.9 (237/792) |
| P Value | Reference | <.001 | <.001 | <.001 | .38 | <.001 |
| RADPEER Score 2b or 3b, % | 37.4 (296/792) | 21.8 (173/792) | 63.5 (503/792) | 42.4 (336/792) | 29.9 (237/792) | 43.2 (342/792) |
| P Value | Reference | <.001 | <.001 | .02 | <.001 | .008 |
| Clinically Acceptable Reports by Standard Criterion, % | 69.7 (552/792) | 85.0 (673/792) | 34.0 (269/792) | 58.6 (464/792) | 68.2 (540/792) | 58.6 (464/792) |
| P Value | Reference | <.001 | <.001 | <.001 | .45 | <.001 |
| Clinically Acceptable Reports by Stringent Criterion, % | 29.8 (236/792) | 50.6 (401/792) | 14.8 (117/792) | 26.3 (208/792) | 33.7 (267/792) | 29.9 (237/792) |
| P Value | Reference | <.001 | <.001 | .08 | .06 | .95 |
| Hallucination, %* | 0.1 (1/786) | 0.5 (4/786) | 12.5 (98/786) | 21.0 (165/786) | 7.6 (60/786) | 12.8 (101/786) |
| P Value | Reference | .21 | <.001 | <.001 | <.001 | <.001 |
| Clarity of Language, % | 79.2 (627/792) | 84.6 (670/792) | 84.3 (668/792) | 73.4 (581/792) | 67.2 (532/792) | 85.6 (678/792) |
| P Value | Reference | .004 | .006 | .005 | <.001 | .001 |

Clinical acceptability was defined as “acceptable as is” or “with minor revision” (standard), and strictly as “acceptable as is” only (stringent).

Clarity of language was defined as a score of 4 or higher (good to excellent).

P values represent comparisons against the radiologist as the reference.

*Two patients whose original reports contained measurements were excluded from the analysis of hallucinations.

**Supplementary Table 5:** Subgroup Analysis in Patients Younger Than 65 Years

| Variable | Radiologist | AIRead | Lingshu | MAIRA-2 | MedGemma | MedVersa |
|---|---|---|---|---|---|---|
| RADPEER Score 3b, % | 9.3 (60/642) | 4.8 (31/642) | 37.4 (240/642) | 17.0 (109/642) | 13.9 (89/642) | 20.9 (134/642) |
| P Value | Reference | <.001 | <.001 | <.001 | .003 | <.001 |
| RADPEER Score 2b or 3b, % | 21.7 (139/642) | 18.4 (118/642) | 49.2 (316/642) | 29.9 (192/642) | 25.4 (163/642) | 32.6 (209/642) |
| P Value | Reference | .09 | <.001 | <.001 | .06 | <.001 |
| Clinically Acceptable Reports by Standard Criterion, % | 79.9 (513/642) | 84.0 (539/642) | 49.8 (320/642) | 74.3 (477/642) | 75.4 (484/642) | 70.1 (450/642) |
| P Value | Reference | .03 | <.001 | .004 | .02 | <.001 |
| Clinically Acceptable Reports by Stringent Criterion, % | 60.4 (388/642) | 67.4 (433/642) | 38.3 (246/642) | 50.9 (327/642) | 54.2 (348/642) | 51.9 (333/642) |
| P Value | Reference | .001 | <.001 | <.001 | .004 | <.001 |
| Hallucination, %[*] | 0 (0/639) | 0 (0/639) | 9.2 (59/639) | 13.0 (83/639) | 2.7 (17/639) | 11.6 (74/639) |
| P Value | Reference | N/A | N/A | N/A | N/A | N/A |
| Clarity of Language, % | 76.8 (493/642) | 80.8 (519/642) | 92.5 (594/642) | 84.6 (543/642) | 72.9 (468/642) | 91.9 (590/642) |
| P Value | Reference | .07 | <.001 | <.001 | .10 | <.001 |

Clinical acceptability was defined as “acceptable as is” or “with minor revision” (standard), and strictly as “acceptable as is” only (stringent).

Clarity of language was defined as a score of 4 or higher (good to excellent).

P values represent comparisons against the radiologist as the reference.

[*]One patient whose original report included a measurement was excluded from the hallucination analysis.

N/A = not available.

**Supplementary Table 6:** Subgroup Analysis of Chest Radiograph Reports by Resident Radiologists

| **Variable** | **Radiologist** | **AIRead** | **Lingshu** | **MAIRA-2** | **MedGemma** | **MedVersa** |
|---|---|---|---|---|---|---|
| RADPEER Score 3b, % | 13.3 (143/1074) | 5.6 (60/1074) | 43.3 (465/1074) | 24.0 (258/1074) | 15.4 (165/1074) | 25.3 (272/1074) |
| P Value | Reference | <.001 | <.001 | <.001 | .12 | <.001 |
| RADPEER Score 2b or 3b, % | 29.3 (315/1074) | 20.9 (225/1074) | 56.5 (607/1074) | 36.1 (388/1074) | 26.6 (286/1074) | 37.7 (405/1074) |
| P Value | Reference | <.001 | <.001 | <.001 | .11 | <.001 |
| Clinically Acceptable Reports by Standard Criterion, % | 75.1 (807/1074) | 84.4 (906/1074) | 41.4 (445/1074) | 65.9 (708/1074) | 72.9 (783/1074) | 64.5 (693/1074) |
| P Value | Reference | <.001 | <.001 | <.001 | .16 | <.001 |
| Clinically Acceptable Reports by Stringent Criterion, % | 45.3 (487/1074) | 58.3 (626/1074) | 26.0 (279/1074) | 37.2 (400/1074) | 44.8 (481/1074) | 40.5 (435/1074) |
| P Value | Reference | <.001 | <.001 | <.001 | .75 | .005 |
| Hallucination, %* | 0.1 (1/1065) | 0.3 (3/1065) | 11.2 (119/1065) | 16.8 (179/1065) | 3.9 (42/1065) | 12.1 (129/1065) |
| P Value | Reference | .34 | <.001 | <.001 | <.001 | <.001 |
| Clarity of Language, % | 77.8 (836/1074) | 83.1 (892/1074) | 88.5 (951/1074) | 79.2 (851/1074) | 69.8 (750/1074) | 88.5 (951/1074) |
| P Value | Reference | .002 | <.001 | .42 | <.001 | <.001 |

Clinical acceptability was defined as “acceptable as is” or “with minor revision” (standard), and strictly as “acceptable as is” only (stringent).

Clarity of language was defined as a score of 4 or higher (good to excellent).

P values represent comparisons against the radiologist as the reference.

*Three patients whose original reports included a measurement were excluded from the hallucination analysis.

**Supplementary Table 7:** Subgroup Analysis of Chest Radiograph Reports by Board-Certified Radiologists

| Variable | Radiologist | AIRead | Lingshu | MAIRA-2 | MedGemma | MedVersa |
|---|---|---|---|---|---|---|
| RADPEER Score 3b, % | 15.8 (57/360) | 4.4 (16/360) | 41.9 (151/360) | 26.1 (94/360) | 21.1 (76/360) | 27.5 (99/360) |
| P Value | Reference | <.001 | <.001 | <.001 | .04 | <.001 |
| RADPEER Score 2b or 3b, % | 33.3 (120/360) | 18.3 (66/360) | 58.9 (212/360) | 38.9 (140/360) | 31.7 (114/360) | 40.6 (146/360) |
| P Value | Reference | <.001 | <.001 | .07 | .58 | .02 |
| Clinically Acceptable Reports by Standard Criterion, % | 71.7 (258/360) | 85.0 (306/360) | 40.0 (144/360) | 64.7 (233/360) | 66.9 (241/360) | 61.4 (221/360) |
| P Value | Reference | <.001 | <.001 | .02 | .11 | .001 |
| Clinically Acceptable Reports by Stringent Criterion, % | 38.1 (137/360) | 57.8 (208/360) | 23.3 (84/360) | 37.5 (135/360) | 37.2 (134/360) | 37.5 (135/360) |
| P Value | Reference | <.001 | <.001 | .85 | .78 | .85 |
| Hallucination, % | 0 (0/360) | 0.3 (1/360) | 10.6 (38/360) | 19.2 (69/360) | 9.7 (35/360) | 12.8 (46/360) |
| P Value | Reference | N/A | N/A | N/A | N/A | N/A |
| Clarity of Language, % | 78.9 (284/360) | 82.5 (297/360) | 86.4 (311/360) | 75.8 (273/360) | 69.4 (250/360) | 88.1 (317/360) |
| P Value | Reference | .21 | .007 | .32 | .003 | .001 |

Clinical acceptability was defined as “acceptable as is” or “with minor revision” (standard), and strictly as “acceptable as is” only (stringent).

Clarity of language was defined as a score of 4 or higher (good to excellent).

P values represent comparisons against the radiologist as the reference.

N/A = not available.